\documentclass[aps,prb,twocolumn,groupedaddress]{revtex4-2}

\usepackage[T1]{fontenc}

\usepackage{amsmath}
\usepackage{amssymb}
\usepackage{bbm}

\usepackage{xspace}
\usepackage{color}
\usepackage{graphicx}
\usepackage{ulem} 
\usepackage{placeins}

\usepackage[colorlinks=true,linkcolor=blue,citecolor=blue,urlcolor=blue]{hyperref}

\graphicspath{./figures/}

\newcommand*{\imag}{\mathrm{i}}
\newcommand*{\HgTe}{\mbox{HgTe}\xspace}

\newcommand{\commentout}[1]{}
\newcommand{\ie}{\textit{i.e.}\xspace} 	
\newcommand{\eg}{\textit{e.g.}\xspace} 	

\begin{document}


\title{Topological insulator constrictions -- Dirac particles in a magneto-chiral box}




\author{Michael Barth}
\author{Maximilian Fürst}
\author{Raphael Kozlovsky}
\author{Klaus Richter}
\email{klaus.richter@ur.de}
\affiliation{Institut  f\"ur  Theoretische  Physik,  Universit\"at  Regensburg,  93040  Regensburg,  Germany}

\author{Cosimo Gorini}
\email{cosimo.gorini@cea.fr}
\affiliation{Universit\'e Paris-Saclay, CEA, CNRS, SPEC, 91191, Gif-sur-Yvette, France}

%


\date{\today}

\begin{abstract}
We study magneto-transport through topological insulator nanowires shaped in the form of a constriction, as can be obtained by etching techniques. The magnetic field is coaxial, potentially turning the nanowire into a magneto-chiral junction.  We show in a detailed analytical and numerical study that two main transport regimes emerge, depending on the central narrow region being \textit{short} or \textit{long} as compared to the magnetic length at the junction entrance and exit.  In both cases the central region hosts Dirac-particle-in-a-box states due to magnetic confinement, whose conductance properties are strongly influenced by Landau levels at the ends of the constriction. 
Notably, in the low-energy regime only chiral states with a specific handedness can transport charge across the junction.  Based on these properties and general symmetry considerations we argue that the shaped nanowire should exhibit strong magneto-chiral non-reciprocal transport beyond linear response.   We employ a numerical tight-binding implementation of an effective 2D model on a non-homogeneous grid, capable of simulating samples of realistic sizes, and test its soundness against full simulations for scaled-down 3D topological insulator wires. 
\end{abstract}


\maketitle

\section{Introduction}

Three-dimensional topological insulators (3DTIs) are bulk insulators hosting metallic states on their surface.  Thanks to certain global (topological) properties of the Hamiltonians describing such materials, the surface states are robust to various perturbations and have a Dirac-like dispersion characterized by (pseudo)spin-momentum locking \cite{hasan2010}.  3DTI nanostructures can nowadays be fabricated with high precision, and nanowires were realized in different materials \cite{dufouleur2013, safdar2013,hong2014,ziegler2018,behner2023,richter2025}.  Electronic transport across such devices is particularly sensitive to magnetic fields at low temperatures, when the phase coherence length $L_\varphi$ exceeds the nanowire typical geometric size $L$.  In this regime geometric phases of the Aharonov-Bohm family compete with Berry phases intrinsic to the Dirac surface Hamiltonian and with quantum confinement effects \cite{dufouleur2013, safdar2013,hong2014,ziegler2018,behner2023}.  For simple nanowires with (nearly) constant cross-section -- whether circular, rectangular, hexagonal or anything in between -- this was discussed theoretically \cite{bardarson2010, ostrovsky2010, sacksteder2014} and demonstrated experimentally \cite{safdar2013,hong2014,ziegler2018,himmler2023}.  Note that surface magneto-transport signatures are particularly evident in high-quality quasi-ballistic HgTe samples \cite{ziegler2018,himmler2023}, since disordered bulk contributions typical of BiSe-based materials \cite{knispel2017,wang2020} are absent there.

\begin{figure}
	\includegraphics[width=0.8\linewidth]{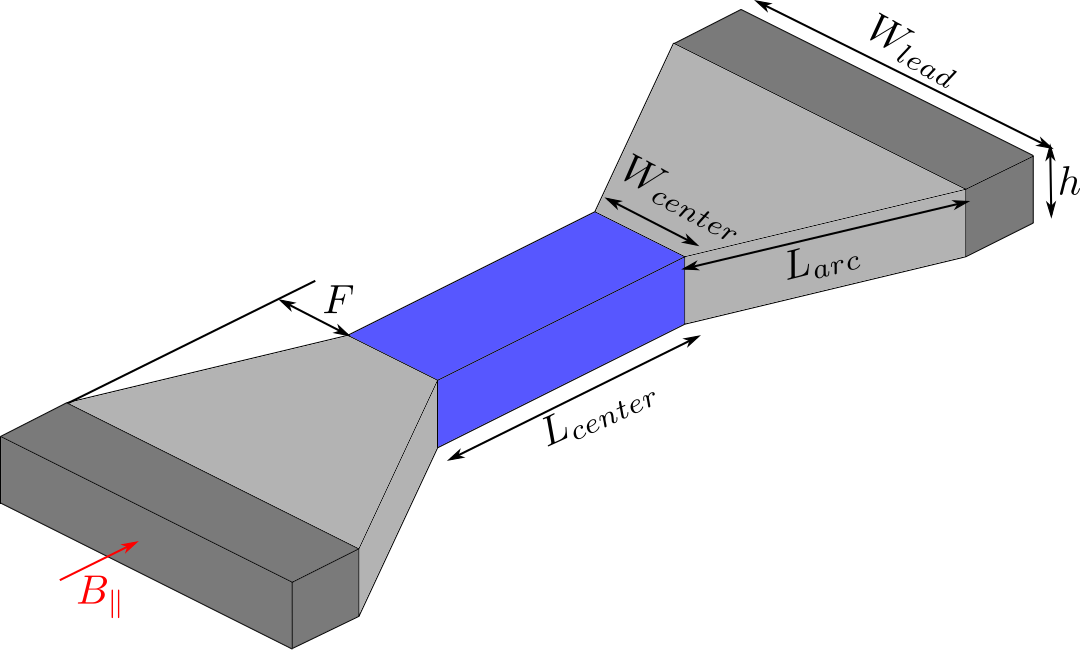}
	\caption{Sketch of a 3DTI nanowire junction consisting of a constriction, \ie a central uniform nanowire connected to two wider TI leads. Such a shape is typical of etched samples.  In order to close the Berry gap of the topological surface states an axial magnetic field is applied to the junction.}
	\label{fig:constriction_schematic}
\end{figure}
The role of the 3DTI nanowire geometry was emphasized recently \cite{kozlovsky2020, graf2020, xypakis2020, ruchi2022, fuerst2024}, starting from the observation that available nanowires typically have a non-constant cross-section, \ie they are {\it shaped}.  Being shaped has indeed a major impact on the nanowire magneto-transport properties, as shown for somewhat idealized, axially-symmetric geometries \cite{kozlovsky2020, graf2020, fuerst2024} not yet realized experimentally.  Complementary to these studies, we here consider instead a more standard shape, sketched in Fig.~\ref{fig:constriction_schematic}, which is realizable notably in etched samples \footnote{For a recent discussion of etching and alternative techniques see \eg \cite{rossler2023} and references therein} and very close to experimentally used devices~\cite{richter2025}. Our main goal is to show how Dirac-type surface transport for such an experimentally available TI-based junction depends on a coaxial magnetic field and can be steered by the latter.

In Sec.~\ref{sec_idea} we will start with a qualitative discussion of magneto-transport in such a junction, grounded on semiclassical and quantum mechanical arguments.  To make these quantitative, we will proceed in Sec.~\ref{sec_model} introducing two models: (i) an effective 2D Hamiltonian for the surface states only, to be discretized in tight-binding form; (ii) a 3D Hamiltonian of BHZ-type, also in tight-binding form.  The 3D model is numerically demanding and will be used to benchmark the 2D one, which allows for simulating micron-size structures.  To this end we will adapt the non-homogeneous tight-binding techniques developed and employed  in Refs.~\cite{kozlovsky2020, graf2020, fuerst2024} to our present geometry.  Our results will be presented in Sec.~\ref{sec_results}, and various aspects thereof will be discussed in Sec.~\ref{sec_chiral_box}.  The emphasis will be put on the chiral properties of the system, and on how the shaped junction could act as a magneto-chiral non-reciprocal conductor, beyond linear response.  We conclude with a brief summary in Sec.~\ref{sec_conclusions}.

\section{The junction: Physical considerations}
\label{sec_idea}

\begin{figure}
    \includegraphics[width=0.8\columnwidth]{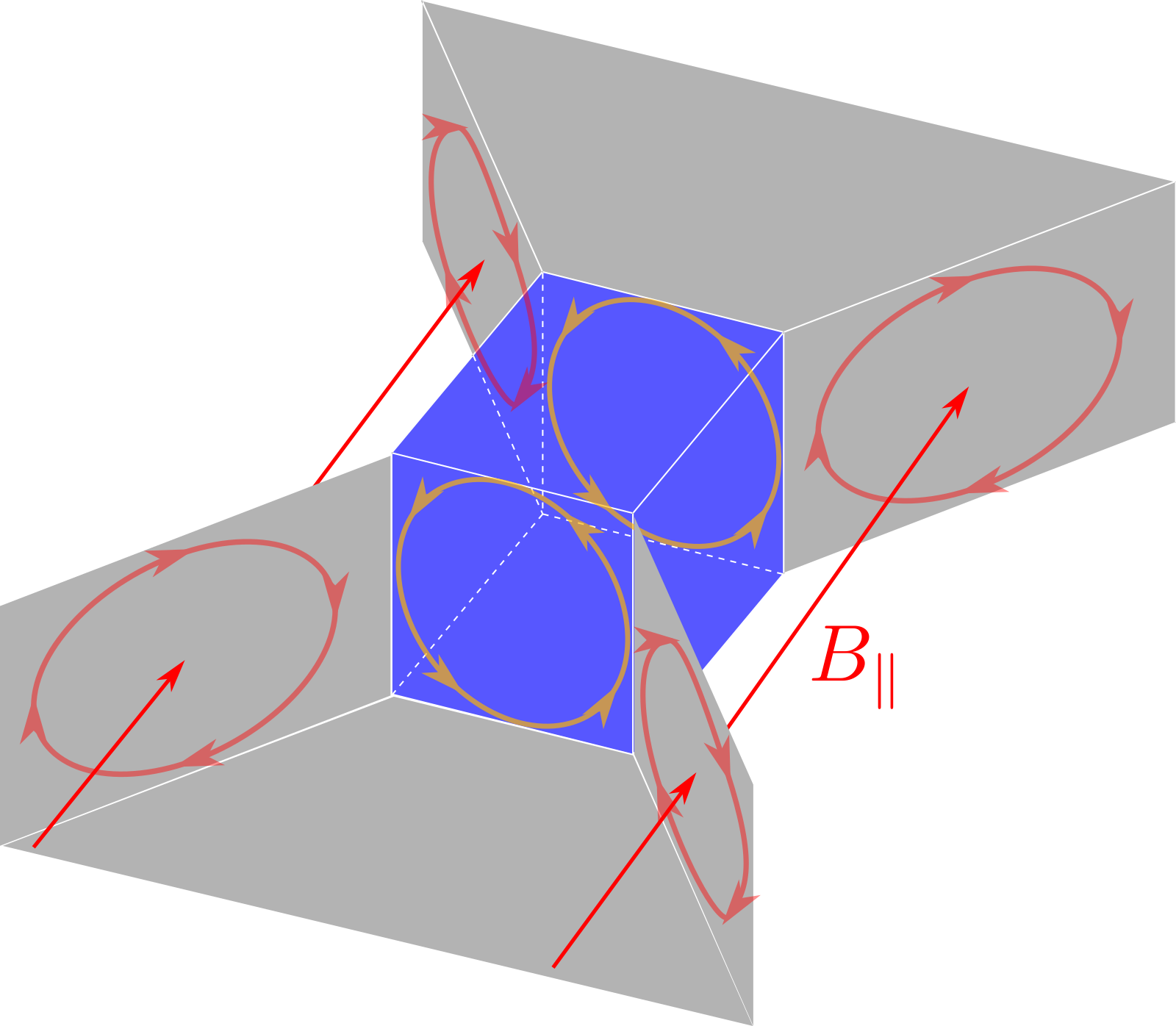}
    \caption{Semiclassical picture for charge carrier dynamics in the chiral Dirac box: surface electronic motion is cyclotron-like on the tapered sides, since the latter are pierced by the $B_\parallel$ field -- recall that for a given handedness of bulk cyclotron motion, the resulting skipping orbits at the edges, shown in red, have opposite handedness.  As a consequence semiclassical trajectories in the central (blue) region also acquire a specific handedness.}
    \label{fig:constriction_chirality}
\end{figure}

A central insight from Refs.~\cite{kozlovsky2020,graf2020} is that Dirac electrons moving along a nanowire of non-constant cross-section in a coaxial magnetic field enclose a varying magnetic flux, and that parts of the conducting surface are pierced by the field.  As a consequence the nanowire can be tuned to act as a purely magnetic trap for Dirac electrons, bypassing the need for electrical gating -- which is furthermore inefficient in Dirac systems due to Klein tunnelling.  The geometry we consider here is however peculiar, since a coaxial magnetic field will have a different qualitative impact in the central region, where the cross-section is constant, and in the tapered entry and exit regions.  Magneto-transport should therefore be determined by the competition between two different electronic behaviours: that in the central region, where the enclosed flux is constant and no field pierces the surfaces, and that in the tapered entry and exit regions, whose sides are pierced by the field and where the enclosed flux varies.  From a semiclassical standpoint one expects the junction to be chiral, as shown in Fig.~\ref{fig:constriction_chirality}: cyclotron motion with a given handedness (red arrows) is imposed on the electrons as they cross the side surfaces pierced by the magnetic field, so that only electrons with one chirality (yellow arrows) will be transmitted.  Clearly, no other classical effect due to the magnetic field can arise from the top and bottom surfaces. 

On the other hand, quantum mechanically the phase mismatch at the entry and exit due to the varying enclosed flux will also play a role, defining a central cavity between two magnetically-tunable barriers -- an effective "Dirac-box''.  Furthermore, in a strong magnetic field the tapered side surface will host Landau Levels, yielding chiral edge channels which are the quantum analog of the red circling arrows.  This suggests that the ``box'' should be chiral, as imposed by the quantum Hall character of the entry/exist side states.  Moreover, its presence could lead to Coulomb-blockade physics of the kind discussed at length in Refs.~\cite{kozlovsky2020,graf2020} \footnote{Standard Coulomb blockade could also arise at $B=0$ if a central cavity forms \eg due to electrostatics or disorder related to etching.}    

We will limit ourselves to considering a fully symmetric junction in the linear response regime, neglecting possible Coulomb effects.  We do this in order to test the validity of the qualitative arguments used so far, aimed at characterising the nature of the ``chiral box''.  We will however argue in the closing that inversion-symmetry breaking, notably in the direction orthogonal to transport, \eg via a bottom gate, would turn the shaped nanowire into a magneto-chiral non-reciprocal junction.  Since non-reciprocal behavior in a two-terminal setup is possible only far from equilibrium, \ie beyond linear response, a dedicated work beyond the scope of the present one is necessary to properly address this interesting aspect.   

\section{Models and numerical implementation}
\label{sec_model}

For the transport calculations we use the tight-binding-based python package Kwant \cite{groth2014}.  We employ two models for the simulation of a constriction in a TI system: (i) a standard effective 2D surface model; (ii) the well-established, full 3D BHZ model.  Option (i) is computationally very efficient and allows us to study experimentally realistic system sizes by capturing the relevant physical properties of the TI surface states.  However the non-trivially shaped surface structure of the TI constriction complicates the implementation, while fermion doubling \cite{susskind1977,nielsen1981,stacey1982} calls for workarounds.  Option (ii) is thus used to check whether the 2D numerical implementation is adequate and that the obtained results reflect the 3D physical setting.


\subsection{Effective 2D surface model}
 
The effective surface Hamiltonian of a cylindrical 3DTI nanowire of radius $R$ in a coaxial magnetic field reads \cite{zhang2010}
\begin{equation}
\label{eq:H-surface_cylinder}
H_{2D} = \hbar v_F \left[ k_z \sigma_z + \left( \frac{1}{R} k_\varphi + \frac{e}{\hbar} A_\varphi\right) \sigma_y \right] - \mu,
\end{equation}
with $\mu$ the chemical potential.  The nanowire axis is along the $z$-direction, $\varphi\in[0,2\pi]$ is the coordinate around the perimeter and $A_\varphi = \frac{\phi}{2\pi R}$ is the vector potential with $\phi$ the flux through the wire cross-section.  To consider non-cylindrical shapes as in Figs.~\ref{fig:constriction_schematic} and \ref{fig:constriction_chirality} we discretize Eq.~\eqref{eq:H-surface_cylinder} on a lattice and write
\begin{equation}
\label{eq:H-surface1}
H_{2D} = \hbar v_F \left[ k_z \sigma_z + \left( \frac{2\pi}{C} k_\varphi + \frac{e}{\hbar} A_\varphi\right) \sigma_y \right] - \mu,
\end{equation}
where $C = 2(W+h)$ is the wire perimeter with width $W$ and height $h$, and $\varphi$ corresponds to the transverse coordinate on the lattice.  The vector potential is $A_\varphi = \phi/C$.  The implementation is numerically straightforward in a real space tight-binding approach: We exchange the wavevectors $k_z, k_\varphi$ by their respective derivative operators, $k_j \rightarrow -i \partial_j, j=z, \varphi$, written in finite-difference form.  One then has
\begin{equation}
\label{eq:H-surface2}
H_{2D} = -i \hbar v_F \left[ \partial_z \sigma_z + \frac{2\pi}{C} \left(\partial_\varphi + i \eta\right) \sigma_y\right] - \mu,
\end{equation}
with $\eta = \phi/\phi_0$ the flux in units of the flux quantum $\phi_0 = h/e$.  The energy spectrum is given by
\begin{equation}\label{eq:Disperion_wire}
E = \pm \hbar v_F \sqrt{k_z^2 + \left(\frac{2\pi}{C} \right)^2 \left(l_n + \eta \right)^2 }.
\end{equation}
Here $l_n = n + 1/2$ is the angular momentum quantum number of the surface states with $n \in {\mathbb Z}$.  The factor $1/2$ stems from the spinor antiperiodic boundary conditions $\Psi(z,\varphi + 2\pi) = \Psi(z,\varphi) \exp\left(-i \pi \right)$.  For the 2D model we fix the Fermi velocity to $v_F = 10^5\,\mathrm{m/s}$.

Putting a Dirac Hamiltonian on a lattice causes fermion doubling \cite{susskind1977,nielsen1981,stacey1982}, whose associated spurious valleys at the edge of the Brillouin zone affect scattering processes.  To circumvent this issue we use the simplest option and add a Wilson mass term 
\begin{equation}\label{Eq:Wilson_Ham}
H_W = \alpha a \hbar v_F \left[ k_z^2 + \left(\frac{2\pi}{C} \right)^2 k_\varphi^2 \right] \sigma_x
\end{equation}
to our Hamiltonian \cite{habib2016,zhou2017}.
Fixing $\alpha = [a/(4\hbar v_F)] E_{\rm gap}$, with $a$ the lattice constant, such a term opens gaps of size $E_{\rm gap}$ at the spurious valleys at the borders of the first Brillouin zone.

To implement the non-trivial nanowire shape from Fig.~\ref{fig:constriction_schematic} we use a non-uniform 2D tight-binding lattice embedded in 3D Cartesian space $(x, y ,z)$.  That is, the coaxial ($z$) and transverse ($\varphi$) lattice constants on our surface change while moving along the wire.  The non-uniformity of the lattice renders the tight-binding Hamiltonian non-Hermitian -- forward and backward hoppings between two given sites are not complex conjugate of each other.  Hermiticity can however be re-established via a local transformation, see Ref.~\cite{kozlovsky2020} for details.  To further describe the magnetic field, taking in particular into account that a component of $B_\parallel$ pierces the tapered constriction sides, we choose the vector potential
\begin{equation}
A_\parallel = B_\parallel 
\begin{pmatrix}
0 \\
x \\
0
\end{pmatrix},
\end{equation}
such that only hoppings along the $y$-direction on the side faces will have a non-trivial phase factor.  The corresponding Peierl's phase reads \cite{kozlovsky-thesis2020}
\begin{align}
\chi_\gamma(\bold{A_\parallel}) &= \exp\left(-i\frac{e}{h} \int_\gamma d\bold{lA_\parallel} \right) \\
&= \exp\left(-i2\pi a_y \frac{B_\parallel C(z)}{2\phi_0} \right),
\end{align}
where the circumference now depends on the $z$-coordinate.  

The Wilson mass term is added inside the leads only, which is sufficient to prevent artificial scattering processes between lead modes.  Finally, to model metallic contacts a high number of open transport modes in the leads is needed.  This is achieved by setting a large lead chemical potential $\mu_L = \mu + \Delta\mu$, with $\Delta\mu$ the offset between the scattering region $\mu$ and the metallic contacts.  The high $\mu_L$ enhances certain transport signatures close to the Dirac point to be discussed below.


\subsection{3D BHZ model}

To double-check that the results obtained via the 2D effective model are sound we also implement full 3DTI constrictions based on the 3D BHZ model.  While the 2D treatment involves a number of numerical tricks (non-uniform grid, anti Fermion doubling terms), the implementation of a 3D model is straightforward.  Its severe drawback is the high numerical cost, which strongly limits the system size.  Nevertheless, already for the accessible system sizes one can observe analogous physical effects, \ie the 3D simulations validate the 2D approach. 

The 3D Hamiltonian is defined as \cite{zhang2009,liu2010} 

\begin{align}\label{eq:H_BHZ}
H^{3d}=&\left[\epsilon({\bf k})-\mu\right]\mathbbm 1_{4\times 4} \nonumber \\
&+
\begin{pmatrix}
M({\bf k})    &  A_1k_z &        0         &           A_2k_- \\
A_1k_z &  -M({\bf k})   &     A_2k_-        &     0           \\
0             &       A_2k_+    &    M({\bf k})    &  -A_1k_z \\
A_2k_+         &        0       &  -A_1k_z  &   -M({\bf k})
\end{pmatrix},
\end{align}
with
\begin{align}
\epsilon({\bf k}) &= C + D_1k_z^2 + D_2(k_x^2+k_y^2), \\
M({\bf k}) &= M - B_1k_z^2 - B_2(k_x^2+k_y^2), \\
k_\pm &= k_x \pm \imag k_y
.
\end{align}
We use simple model parameters which neglect anisotropy effects in the transversal wire directions for better comparison with the 2D results.  The parameters are fixed to $M= 1\,\mathrm{eV}$, $A_1=-1\,\mathrm{eV\,\mathring{A}}$, $A_2=-1\,\mathrm{eV\,\mathring{A}}$, $B_1=1\,\mathrm{eV\,\mathring{A}^2}$, $B_2=1\,\mathrm{eV\,\mathring{A}^2}$, and $C = D_1 = D_2 = 0$.
The external magnetic field is introduced via an appropriate Peierl substitution.

\section{Results: Resonant Charge Transfer}
\label{sec_results}

We show that different transport regimes exist depending on the length of the narrow central region, \ie whether this is \textit{long} or \textit{short}.  It will be manifest below that a junction behaves as a \textit{short} or \textit{long} one depending on whether the length of the narrow region is \textit{shorter} or \textit{longer} than the magnetic length $l_B = \sqrt{\hbar/e\tilde{B}_\parallel}$.  Here $\tilde{B}_\parallel$ is the component of $B_\parallel$ orthogonal to the tapered side faces connecting the leads to the central region.  Notice that such a length exists only at the entrance/exit to the central narrow segment, not within it -- in the central segment the field is parallel to the surface, so that $l_B \to \infty$, no matter how long or short the junction may be.  Nevertheless, it is the role of $l_B$ that fundamentally sets the transport character of the 3DTI shaped junction, in a way that will be discussed in detail below.  We study the problem via the effective surface model, and then check the results by additional 3D calculations. 
\begin{figure}
	\centering
	\includegraphics[width=\columnwidth]{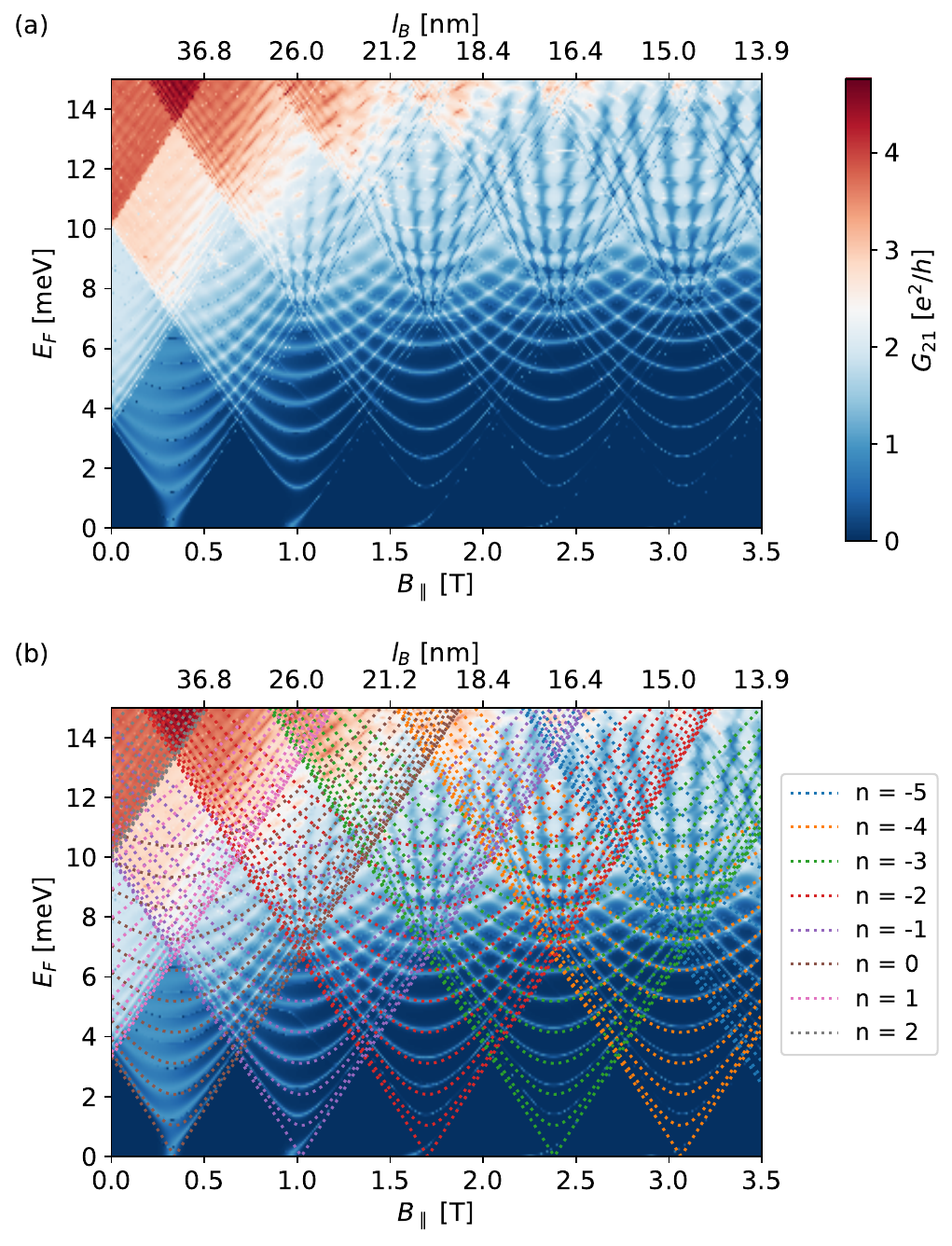}
	\caption{(a) Color plot of the conductance as function of the applied magnetic field $B_\parallel$ and the Fermi energy $E_F$. (b) Energy levels from Eq.~\eqref{eq:Disperion_resonant_states} (dashed lines) superimposed over the conductance map shown in (a). Different colors of the dashed lines correspond to different angular momentum quantum numbers $n$ while different branches of the same color correspond to different longitudinal momentum quantum numbers $m$. The lowest linear branch stems from the $m=0$ state.
     The system parameters are fixed to the following values: $H=80\,\mathrm{nm}$, $W_{lead}=150\,\mathrm{nm}$, $W_{center}=75\,\mathrm{nm}$, $L_{arc}=150\,\mathrm{nm}$, $L_{center}=1000\,\mathrm{nm}$, $\Delta\mu=30\,\mathrm{meV}, \mu=0 \mathrm{meV}$.
    }
	\label{fig:2d_conductance_L1000nm}
\end{figure}

\subsection{2D model: long central segment}
\label{subsec_2D_long}

For the system shown in Fig.~\ref{fig:constriction_schematic} the geometric parameters are chosen so as to be in an experimentally feasible range. 
We fix the parameters of our 2D simulations to $H=80\,\mathrm{nm}$, $W_{lead}=150\,\mathrm{nm}$, $W_{center}=75\,\mathrm{nm}$, $L_{arc}=150\,\mathrm{nm}$, $\mu=0$ and $\Delta\mu=30\,\mathrm{meV}$ and use these values throughout the following subsections. 
The critical parameter to be varied is the length $L_{center}$ of the central segment.  In this subsection we take $L_{center}=1000\,\mathrm{nm}$.

\begin{figure}
	\centering
	\includegraphics[width=\columnwidth]{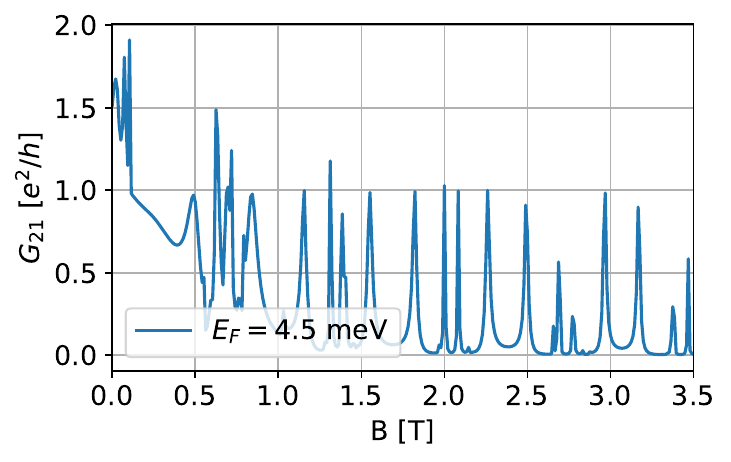}
	\caption{Conductance as a function of the applied magnetic field $B_\parallel$ and the Fermi energy $E_F=4.5\mathrm{meV}$. The curve corresponds to a constant energy line cut through Fig.~\ref{fig:2d_conductance_L1000nm}~(a).}
	\label{fig:2d_conductance_linecut_L1000nm}
\end{figure}

Fig.~\ref{fig:2d_conductance_L1000nm}~(a) shows the conductance of the 3DTI constriction as a function of the magnetic field and the Fermi energy $E_F$ within the junction.  Notice that $L_{center} \gg l_B$ in the full magnetic field range.  The emergence of regularly-spaced conductance traces of hyperbolic shape from a certain magnetic field strength on is evident.  Sets of such traces appear periodically with respect to the magnetic field, all equally spaced as the Fermi energy increases.  The periodicity of the conductance line structures is given by the magnetic field that is necessary to thread one magnetic flux quantum through the cross-section of the narrow central segment, \ie $B=0.7\,\mathrm{T}$ in our case.  The hyperbolic conductance traces are furthemore approximately quantized to integer multiples of one conductance quantum $\frac{e^2}{h}$ from a certain field strength on, see Fig.~\ref{fig:2d_conductance_linecut_L1000nm}.

These features can be explained by a simple quantum mechanical ``Dirac particle in a box'' picture.  The magnetic field component that is perpendicular to the side tapered faces at the constriction entrance/exit leads to the formation of local quantum Hall states.  The latter act as magnetic barriers \cite{kozlovsky2020, graf2020}, trapping the Dirac electrons within the central region.  Therein the continuous momentum $k_z$ of Eq.~\eqref{eq:Disperion_wire} is quantized, yielding resonant levels.  The simplest approximation is to assume the electrons are strictly trapped in the central ``box'', setting the longitudinal momentum to $k_z = m \pi/ L_{center}$ with $m \in {\mathbb Z}$.  The resonant energies are then
\begin{equation}\label{eq:Disperion_resonant_states}
E = \pm \hbar v_F \sqrt{\left(m\frac{\pi}{L_{center}}\right)^2 + \left(\frac{2\pi}{C} \right)^2 \left(l_n + \eta \right)^2 },
\end{equation}
where the only continuous quantity is the flux ratio $\eta$.

Using Eq.~\eqref{eq:Disperion_resonant_states} we can compare this resonant spectrum to the numerical conductance traces, as shown by superposition in Fig.~\ref{fig:2d_conductance_L1000nm}~(b).  For every angular momentum quantum number $n$ one gets a periodically repeated structure, visualized with differently colored dashed lines.  Furthermore, for each such periodic set, the quantized conductance traces arise from different longitudinal momentum quantum numbers $m$.  
Once the magnetic field strength is beyond a certain threshold, which is determined by comparing the magnetic length $l_B$ with the relevant geometrical scale of the sample, the conductance is fully quantized~\footnote{The values of $l_B$ can be found on the upper horizontal axis of most plots.}.  The $m=0$ state, which strictly speaking describes a circulating mode with zero longitudinal kinetic energy, corresponds to the lowest branch in each set.  For Fermi energies close to the Dirac point the lowest branches bend smoothly towards $E_F=0$.  This behaviour would hardly be visible without doped leads: the $m=0$ states do not have kinetic energy to overcome the magnetic barriers at the constriction entrance/exit, and this is then provided by the incoming lead modes with higher energy $\Delta\mu$.  The magnetic barriers indeed appear when the magnetic length $l_B$ becomes shorter than the height and the length $L_{arc}$ of the faces, ensuring the formation of Landau levels on the tapered side faces.

\begin{figure}
	\centering
	\includegraphics[width=\columnwidth]{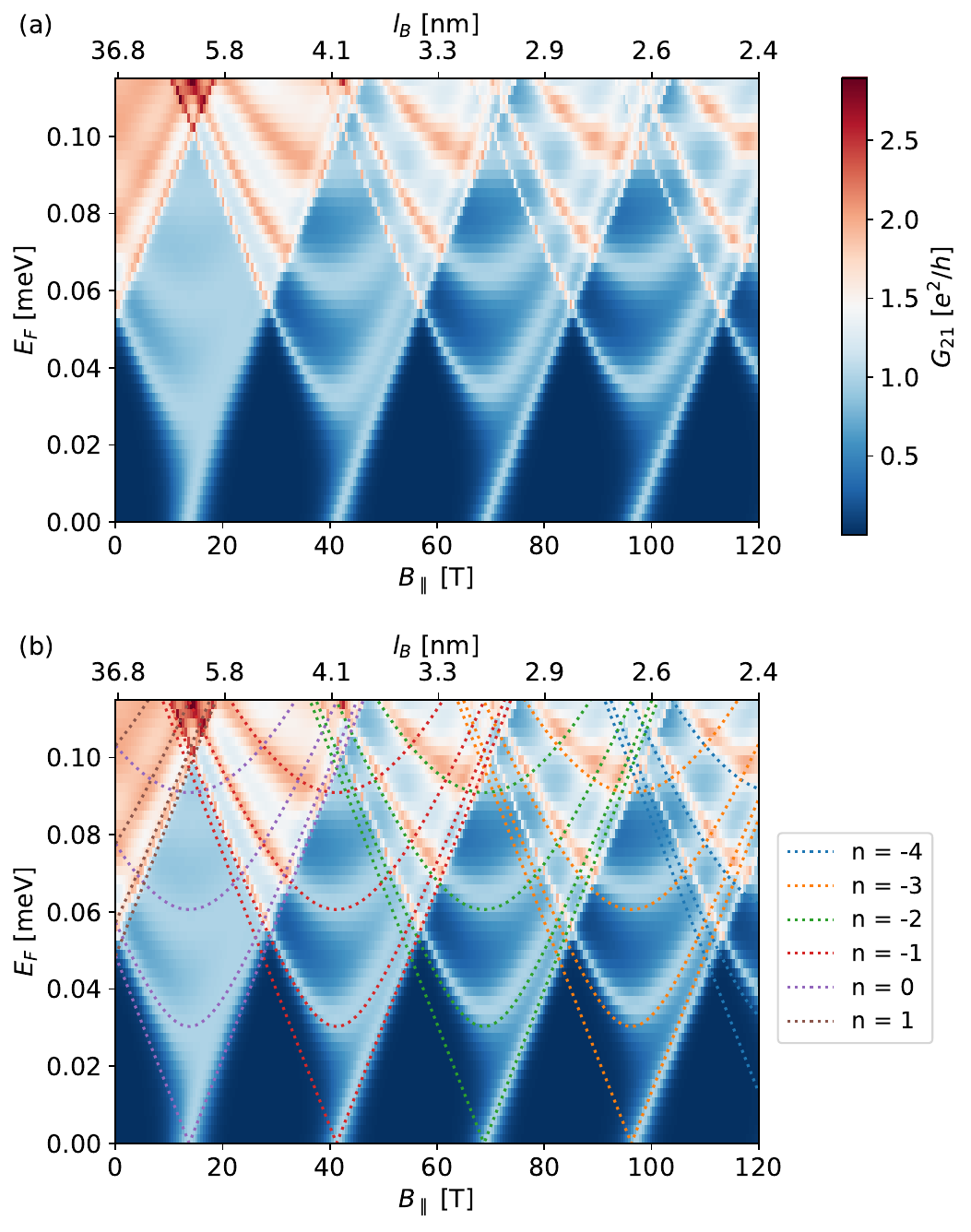}
	\caption{(a) Conductance as a function of magnetic field $B_\parallel$ and Fermi energy $E_F$, obtained from the 3D model of Eq.~\eqref{eq:H_BHZ}. (b) Superimposed energies from  Eq.~\eqref{eq:Disperion_resonant_states} (dashed lines) with $\eta_{eff}$ onto the conductance map from (a). As in Fig.~\ref{fig:2d_conductance_L1000nm}, different dashed line colors correspond to different angular momentum quantum numbers $n$, while different branches of the same color correspond to different longitudinal momentum quantum numbers $m$. The system parameters are: $a=1\,\mathrm{nm}$, $H=20\,\mathrm{nm}$, $W_{lead}=30\,\mathrm{nm}$, $W_{center}=10\,\mathrm{nm}$, $L_{arc}=10\,\mathrm{nm}$, $L_{center}=100\,\mathrm{nm}$, $\Delta\mu=0.38\,\mathrm{meV}$. }
	\label{fig:3d_conductance_L100}
\end{figure}

\subsection{3D model: long central segment}
\label{subsec_3D_long}

We use 3D simulations based on Eq.~\eqref{eq:H_BHZ} to test the physical picture obtained from the effective 2D model.  The number of lattice points is now limited, as the computational cost rises very fast with system size.  The parameters are thus fixed to $a=1\,\mathrm{nm}$ (lattice constant for the cubic grid), $H=20\,\mathrm{nm}$, $W_{lead}=30\,\mathrm{nm}$, $W_{center}=10\,\mathrm{nm}$, $L_{arc}=10\,\mathrm{nm}$, $L_{center}=100\,\mathrm{nm}$ and $\Delta\mu=0.38\,\mathrm{meV}$ to reach a compromise between computational costs and reasonable length scales.  Our parameter choice results in a \textit{hard} constriction without any tapered sides, where the narrow central wire segment is directly coupled to wider leads.  This is done to reduce the field range required to induce magnetic confinement, as the full $B_\parallel$ pierces the (now flat) sides.  

The conductance of the 3D constriction is again calculated as a function of the magnetic field and of the Fermi energy within the junction, and is shown in Fig.~\ref{fig:3d_conductance_L100}~(a).  As in the 2D model, we can observe the emergence of quantized conductance lines which are equally spaced in energy and periodic in the magnetic field.  The quantization is more pronounced when the magnetic length becomes comparable to the length scales of the (flat) side faces.  Note that the necessary field strengths are much larger due to the small wire cross section.  Such values are clearly not realistic; They only serve the purpose of showing that the same physics as in the 2D model is observed given similar ratios between $l_B$ and the lengths defining the nanowire geometry.  

To compare the energies from  Eq.~\eqref{eq:Disperion_resonant_states} with the corresponding conductance map one must account for the finite penetration depth $\lambda_s$ of the surface states into the bulk of the material.  For the actual model parameters we assume a penetration depth of $\lambda_s = 1\,\mathrm{nm}$ and rescale the flux ratio according to 
\begin{equation}
\eta_{eff} = \frac{\phi_{eff}}{\phi_0} = \frac{B_\parallel (W_{center}-\lambda_s)(H-\lambda_s)}{\phi_0}.
\end{equation} 
Given $\eta_{eff}$ and the effective central length of $L_{center}\approx85~\mathrm{nm}$, a good agreement between Eq.~\eqref{eq:Disperion_resonant_states} and the quantized conductance traces is obtained, see Fig.~\ref{fig:3d_conductance_L100}~(b).  This confirms that the effective 2D implementation from the previous subsection captures and reproduces the relevant physics of the shaped nanowire.


\subsection{2D model: short central segment}
\label{subsec_2D_short}

Let us now shorten the box, \ie the narrow central segment of the junction.  Eq.~\eqref{eq:Disperion_resonant_states} suggests that reducing $L_{center}$ will increase the spacing between resonant energy levels within the box.  Naively one expects that below a certain length only the $m=0$ state with its linear flux dependence will be present in the low energy regime.  

\begin{figure}
	\centering
	\includegraphics[width=\columnwidth]{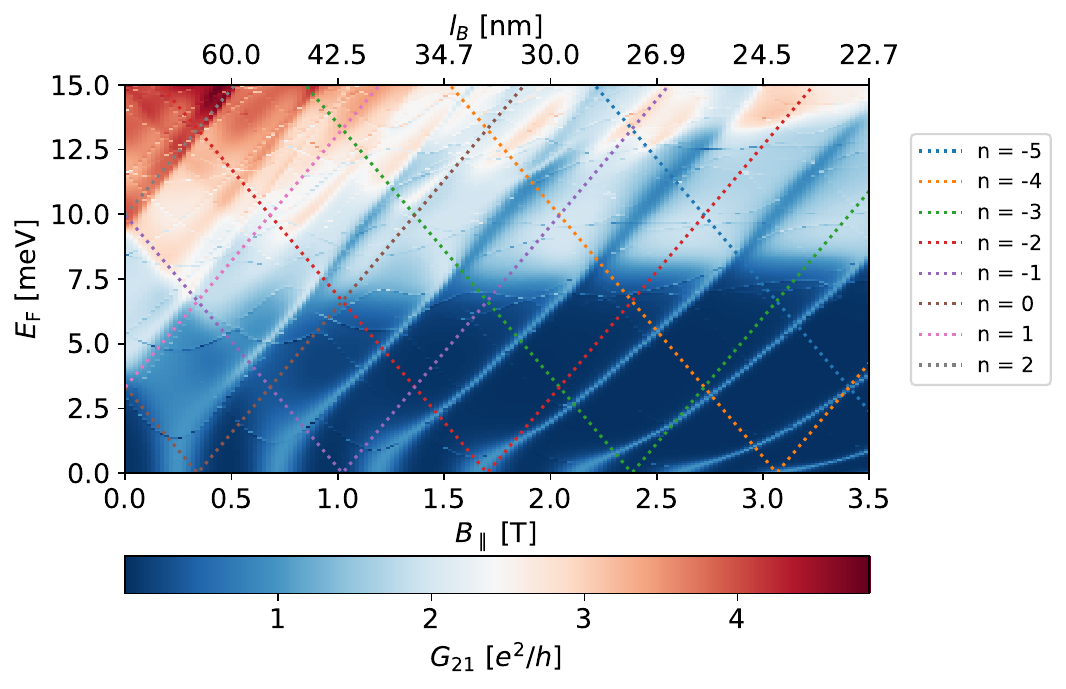}
	\caption{Conductance as a function of applied magnetic field $B_\parallel$ and Fermi energy $E_F$ in the short junction limit. The system parameters are: $H=80\,\mathrm{nm}$, $W_{lead}=150\,\mathrm{nm}$, $W_{center}=75\,\mathrm{nm}$, $L_{arc}=150\,\mathrm{nm}$, $L_{center}=10\,\mathrm{nm}$, $\Delta\mu=30\,\mathrm{meV}$.}
	\label{fig:2d_conductance_L10nm_fit}
\end{figure}

For the short-junction simulation we keep all system parameters as before except for the length, which is shortened to $L_{center}=10\,\mathrm{nm}$.  This means that $L_{center} \lesssim l_B$ in the full magnetic field range.  Figure~\ref{fig:2d_conductance_L10nm_fit} shows the conductance map with superimposed resonant energy levels obtained from Eq.~\eqref{eq:Disperion_resonant_states}.  Clearly, the simple particle in a box picture does not quite work in this regime.  If, as expected, only single energy levels appear periodically in the low energy range, they strongly deviate from a linear flux dependence.  For stronger magnetic fields the conductance lines smoothly bend while approaching the Dirac point at $E_F=0$.  This feature signals the emergence of Landau levels on the tapered side faces, and specifically in this case the zero-energy Landau level: the short-junction behavior seems to be dominated by the two constricting regions, rather than by the states trapped in the central box as in Subsecs.~\ref{subsec_2D_long}, \ref{subsec_3D_long}.


\begin{figure}
	\centering
	\includegraphics[width=\columnwidth]{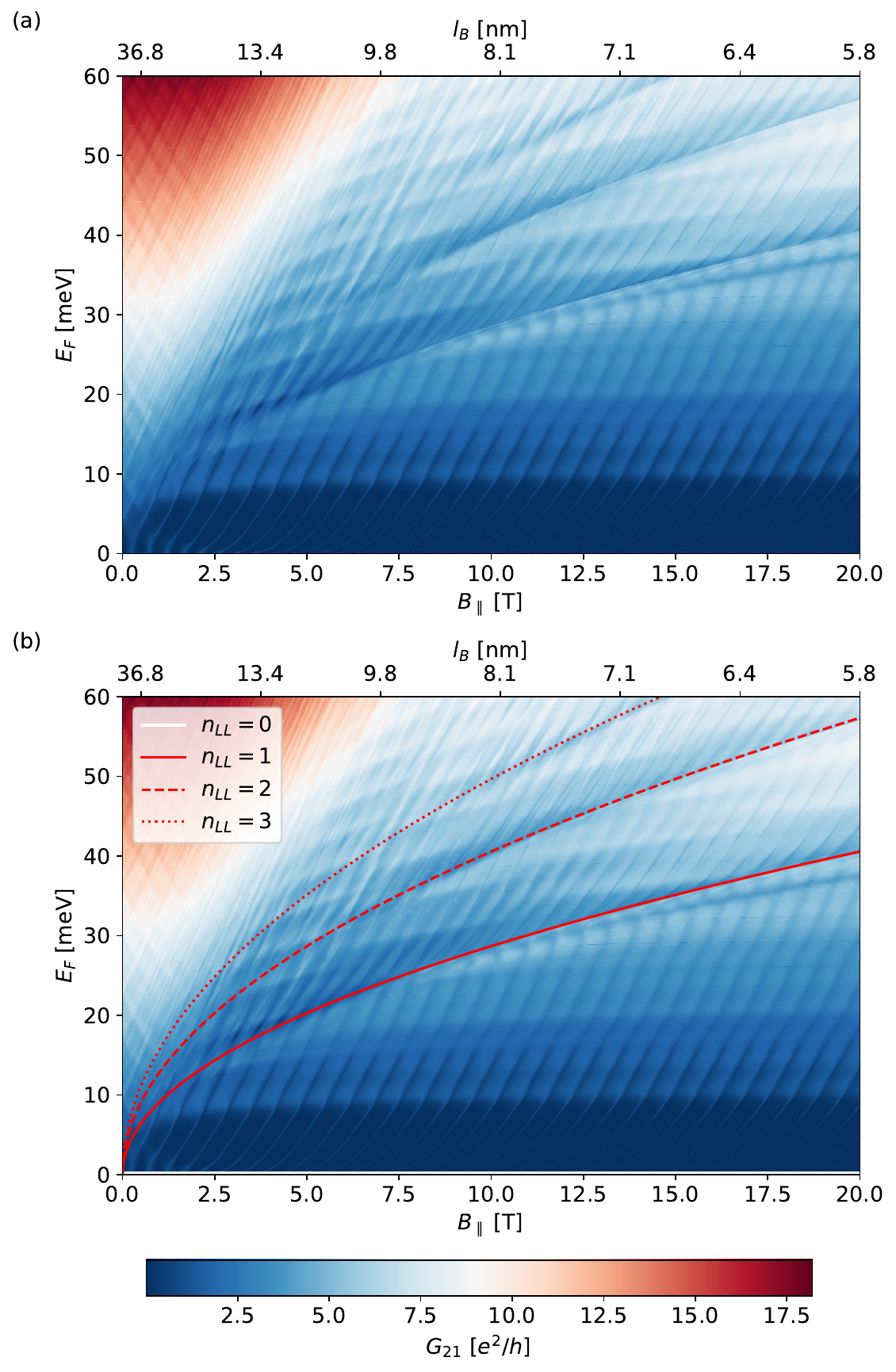}
	\caption{(a) Conductance map for the same system parameters as those  used for Fig.~\ref{fig:2d_conductance_L10nm_fit}. The energy as well as the magnetic field range were increased to larger magnitudes. The conductance lines are clearly overlapped with a larger line structure corresponding to the development of Landau levels. (b) Comparison of the $n= 0,1,2\,\mathrm{and}\,3$ Landau levels calculated from Eq.~\eqref{eq:Dirac_LL} and the conductance map of (a).  Fainter Landau level-like traces -- not highlighted -- are spurious effects due to the Wilson mass term, Eq.~\eqref{Eq:Wilson_Ham}, which artificially splits the lead modes.}
	\label{fig:2d_conductance_L10nm_large_parameter_range}
\end{figure}

To test this hypothesis and visualize the appearance of quantum Hall states on the side faces we increase the energy and magnetic field range of the transport simulations.  Figure~\ref{fig:2d_conductance_L10nm_large_parameter_range}~(a) shows the same conductance map as in Fig.~\ref{fig:2d_conductance_L10nm_fit} but for extended $B_\parallel$ and $E_F$ ranges.  The map is overlaid with a widely-spaced rising line structure, to which the underlying resonant conductance traces smoothly merge.  Such a structure is the direct imprint of Dirac Landau levels located on the tapered side faces at the junction entrance/exit.
%
%

To see this, recall that Dirac Landau level energies read \cite{yin2017}
\begin{equation}\label{eq:Dirac_LL}
    E_{LL} = \sqrt{2e\hbar n_{LL}}v_F\sqrt{B_\parallel \frac{F}{L_{arc}}}, 
\end{equation}
with $n_{LL} \in {\mathbb Z}$ the Landau level principal quantum number, and $B_\parallel F/L_{arc} \equiv \tilde{B}_\parallel$ the field component orthogonal to the slanted sides.  Plotting such energy levels for the system parameters of our simulations, a perfect fit with the widely-spaced conductance features is obtained, see Fig.~\ref{fig:2d_conductance_L10nm_large_parameter_range}(b).
Transport across the junction is indeed dominated by the condensation of Landau levels at the entrance/exit to the central box.  That is, in the short-junction regime $L_{center} \lesssim l_B$ the central segment acts less as a trap (box) and more as a barrier between states at its far ends, while the latter become more and more Landau level-like as $B_\parallel$ increases.  Note that on the highlighted Landau level lines the conductance drops, \ie these states tend to pin charge carriers on their side of the junction.  On the other hand, in a Landau gap charge carriers bypass the gapped sides and propagate via the ungapped top and bottom surfaces yielding a higher conductance. 

\subsection{3d model: short central segment}
\label{subsec_3D_short}

\begin{figure}
	\centering
	\includegraphics[width=\columnwidth]{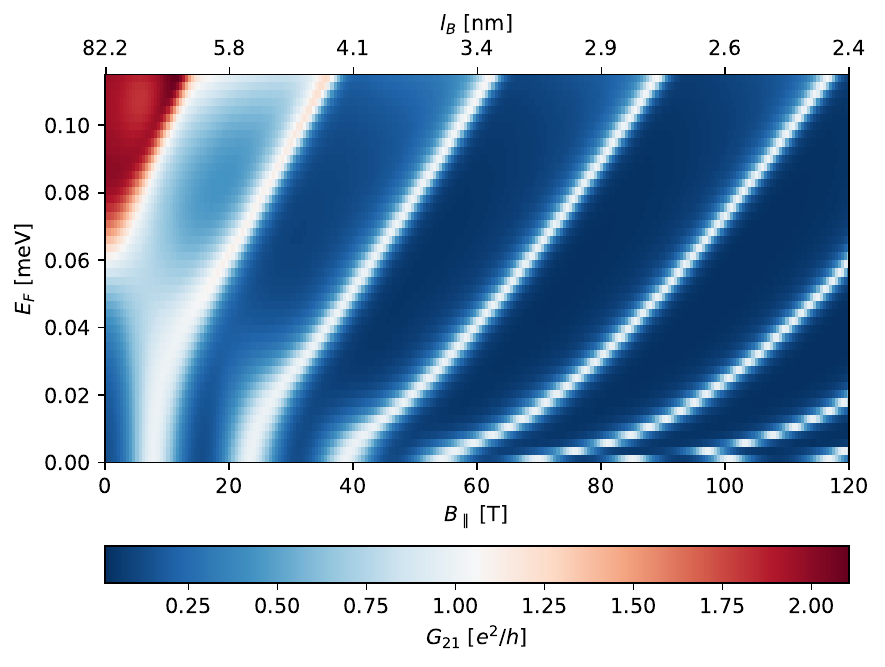}
	\caption{Conductance map as a function of the Fermi energy $E_F$ and of the magnetic field $B_\parallel$ obtained from the 3D model for a short central segment with length $L_{center}=10~\mathrm{nm}$. The central junction is smoothly connected to the wider leads by contracting regions with $L_{arc}=18~\mathrm{nm}$ and $F=10~\mathrm{nm}$. As in the effective 2D implementation, only conductance traces with positive slope appear, and the latter smoothly bend towards $E_F=0$.}
	\label{fig:3d_short_conductance_map}
\end{figure}

As in previous subsections, we test the conclusions obtained from the effective 2D model against the results of a 3D simulation in a downscaled system.  The central region being now considerably shorter has an advantage: the overall number of lattice sites is low enough to let us perform 3D simulations including tapered entrance/exit. (Recall that in Subsec.~\ref{subsec_3D_long} the limitation on lattice sites forced us to consider a \textit{hard} junction, \ie with flat rather than tapered connections to the leads).  We now set $L_{arc}=18~\mathrm{nm}$ and $F=10~\mathrm{nm}$.  The narrow central segment length is $L_{center}=10~\mathrm{nm}$, while all other parameters remains as in Subsec.~\ref{subsec_3D_long}.  Figure~\ref{fig:3d_short_conductance_map} shows the conductance plot at low energies, for a magnetic field range where $l_B$ is still larger than our lattice constant, such that Peierl's substitution remains reasonable.  

As for the 2D simulations, for increasing magnetic field strength resonant conductance lines have a positive slope and smoothly bend towards the zero-energy axis: transport is dominated by the zeroth Landau level emerging at the constriction entrance/exit.
%

\begin{figure}
	\centering
	\includegraphics[width=\columnwidth]{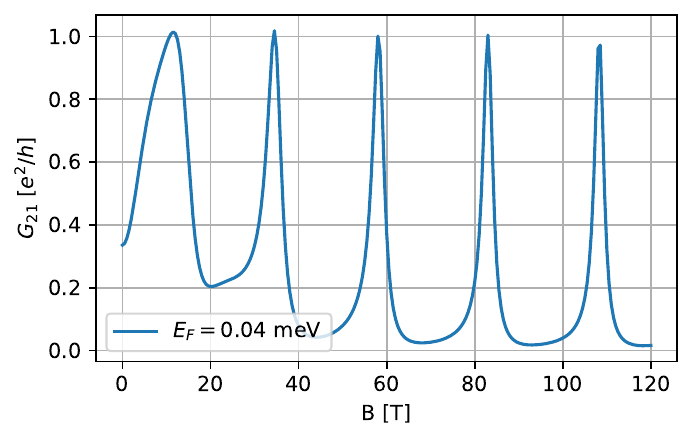}
	\caption{Conductance as function of the coaxial magnetic field at fixed Fermi energy $E_F=0.04~\mathrm{meV}$, computed for a short constriction via a 3D model. The curve corresponds to a line cut through the conductance map of Fig.~\ref{fig:3d_short_conductance_map}. Resonant conductance peaks which are quantized to roughly one conductance quantum are periodic in the magnetic field.}
	\label{fig:3d_conductance_line_L40}
\end{figure}

Note also that the conductance peaks are quantized to roughly one conductance quantum bewond a certain field strength.  To see this more explicitly a cut at constant energy $E_F=0.04~\mathrm{meV}$ through the conductance map is shown in Fig.~\ref{fig:3d_conductance_line_L40}.  At low field the magnetic barriers are too weak to suppress tunneling through the junction, so that peaks are broader.  Their resonant character grows with growing field strength.  

The 3D results confirms once more the physical picture obtained for realistically-sized systems via our effective 2D treatment.

\begin{figure*}
	\centering
	\includegraphics[width=0.7\linewidth]{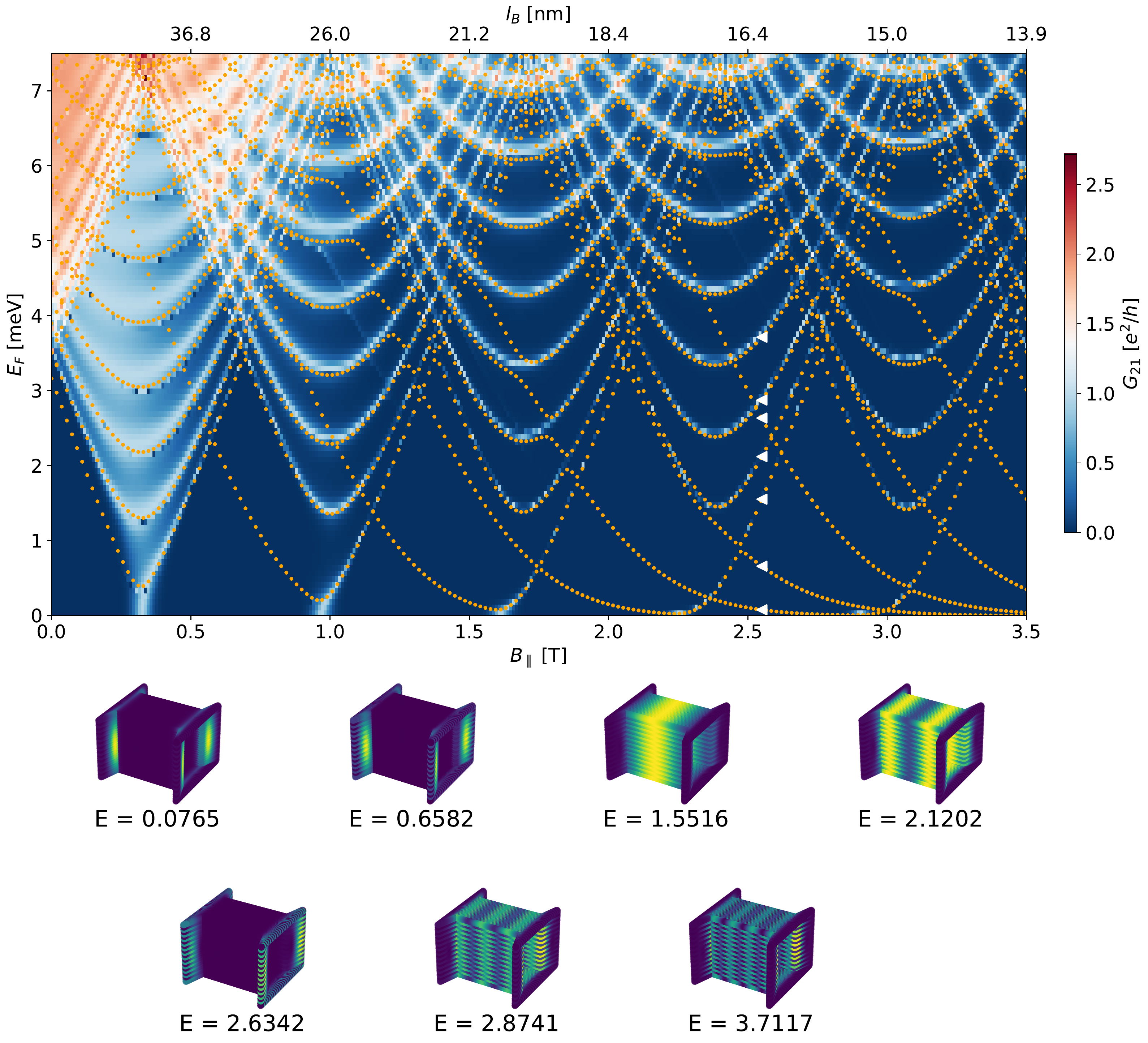}
	\caption{Comparison between the conductance map of a long junction corresponding to the system of Fig.~\ref{fig:2d_conductance_L1000nm} and the numerically calculated eigenspectrum of the finite constriction region. The latter is shown in orange dotted lines. It fits perfectly with the observed conductance features, except for the lowest negative-slope branches to which vanishing conductance is associated.  This can be understood looking at the probability densities of the seven lowest eigenstates -- marked as white triangles -- which are plotted at a fixed magnetic field strength of $B_\parallel=2.55~\mathrm{T}$.  States with $m=0$ and a negative slope are localized at the side faces and thus cannot contribute to transport.  On the contrary, states belonging to positive-slope branches are extended throughout the junction and thus conduct.  They correspond to resonant Dirac states in the central wire segment.}
	\label{fig:2d_conductance_long_junction_with_eigenenergies_and_densities}
\end{figure*}

\section{Low-energy magneto-chiral Dirac box}
\label{sec_chiral_box}

\subsection{Structure of resonant states}

In both long and short junctions we found quantized conductance signatures stemming from resonant energy levels in the central wire segment.  Depending on the system specifics these ``particle in a box'' signatures may be completely dominant -- long junctions -- or heavily distorted by the Landau levels in the regions connecting the former to the leads -- short junctions.  Independently of such differences, and for both long and short wires, there is one crucial aspect common to all conductance maps: There is no transport signature of the $m=0$ state with negative slope.   
This is an explicit quantum mechanical signature of the chirality of the shaped 3DTI junction, and confirms the intuition from the semiclassical argument introduced in Sec.~\ref{sec_idea}, see Fig.~\ref{fig:constriction_chirality}.  To see this, at first qualitatively, recall that the $m=0$ state has zero longitudinal momentum, \textit{ergo} its energy is purely given by angular momentum:
\begin{equation}
    E_0 = \pm \hbar v_F \frac{2\pi}{C} |n+0.5+\eta|.
\end{equation}

The analogy with the eigenstates of an Aharonov-Bohm ring 
\cite{aharonov1959} for a Dirac system~\cite{Wurm2010} 
is manifest, and thus with the associated and well-known persistent currents \cite{buttiker1986,cheung1988,eckern1995,Richter96,crepin2016}.  Such currents are proportional to the derivative of the energy levels with respect to the flux, which in our case reads
\begin{equation}
    I_\mp = - \frac{\partial E_0}{\partial \phi} = \mp \hbar v_\mathrm{F} \frac{2\pi}{C_{center}}\frac{1}{\phi_0} = \mp \frac{ev_F}{C_{center}}.
\end{equation}
The Aharonov-Bohm analogy stops when one considers that the central region states, even those at $m=0$, may also extend in the longitudinal direction $z$ -- we are dealing with a tube, not with a ring.  If states sustaining both $I_+$ and $I_-$ currents should exist, in our system only one species, say the one yielding $I_+$, is extended along $z$ and can thus transport charge to the other side of the junction.  This is due to the quantum Hall states emerging on both sides of the narrow central region: the latter must be mode-matched to those in the central region, and the matching will select one chirality as the only propagating one.  On the other hand, states with higher momentum $m\neq0$ have finite longitudinal kinetic energy and can thus propagate along the wire length with less concern for matching effects. 
%

To verify this qualitative picture we numerically compute the constriction eigenenergies and eigenstates.  The spectrum is computed by diagonalization of a finite tight-binding system consisting of the central segment and the two tapered regions, shown in blue and light grey, respectively, in Fig.~\ref{fig:constriction_schematic}.  We take the long-junction geometry of Sec.~\ref{subsec_2D_long}, and also directly check the quantized particle in a box picture.  The resulting eigenenergies are shown as orange dots in Fig.~\ref{fig:2d_conductance_long_junction_with_eigenenergies_and_densities}, superimposed on the corresponding conductance map.  Most values lie right on top of the resonant conductance lines.  However, there is also an ensemble of energy levels in regions of zero conductance, \ie eigenenergies to which no conductance signature can be assigned.  Such energies belong to the $m=0$ states with negative slope for all $n$ angular momentum quantum numbers.  These states bend towards $E_F=0$ for increasing magnetic field, showing the influence of the emerging zeroth Landau levels on the side faces.  This influence can be seen explicitly in the bottom panel of Fig.~\ref{fig:2d_conductance_long_junction_with_eigenenergies_and_densities}, from a plot of the probability density of the seven lowest eigenstates at a fixed field $B_\parallel=2.55~\mathrm{T}$. This value is chosen such that magnetic confinement is strong enough to develop resonant energies in the junction.  The densities are labeled by their corresponding energy values, and their position on the conductance maps is marked by white triangles.  The two lowest energy states belong to different angular momentum quantum numbers $n \neq n'$, but both share the same longitudinal quantum number $m=0$.  They are strongly localized at the side faces of the contracting regions and thus cannot contribute to transport -- this is another aspect of the pinning effect of Landau levels. 
In contrast, the third state in the energy sequence extends throughout the central wire segment, despite the fact that it has zero longitudinal momentum $m=0$ just as the two previous ones.  This is due to the different handedness of the associated current, as seen from the positive slope of the energy branch to which it belongs.
The fourth state has $m=1$ and therefore two maxima in the local density of states, as expected from the particle in a box picture.  The fifth state is again a non-conducting state with $m=0$ localized at the junction edges.  The sixth and seventh have $m=2$ and $m=3$, respectively, and belong to positive-slope branches.  As expected, they are both conducting and yield a local density of states with an increasing number of maxima and nodes, further validating the particle in a box picture.

\subsection{Non-reciprocal transport}
\label{subsec_non-reciprocal}

Consider the general expression for the current in a two-terminal device, written as a power series in the applied bias $V$
\begin{equation}
\label{eq_current}
I(V,B) = G_0(B) V + G_1(B)V^2 + \dots.
\end{equation}
The linear response magneto-conductance $G_0$ is strongly constrained by Onsager-Casimir reciprocity \cite{onsager1931, casimir1945}.  In particular, this object must be even in the applied magnetic field, $G_0(B) = G_0(-B)$ -- this is the case for our junction.  Indeed, we do not show conductance traces at negative $B_\parallel$ as they are mirror-images of the positive-field ones.  Furthermore, it is obvious that the linear response current changes sign reversing the bias $I(V) = -I(-V)$.   

In the non-linear regime non-reciprocal two-terminal transport becomes accessible \cite{christen1996, sanchez2004, lofgren2004, angers2007}, which in particular means that the current magnitude in general depends on the bias direction, $I(V) \neq -I(-V)$.  To leading order one thus has $G_1 \neq 0$.  In the absence of a magnetic field a requirement is that inversion symmetry along the conductor is broken.  What if a finite $B$ is present?  In this case even if longitudinal inversion symmetry is preserved non-reciprocity is possible, with a linear-in-$B$ behavior $G_1(B) \neq G_1(-B)$ appearing if inversion symmetry is broken in the direction transverse to transport \cite{sanchez2004, lofgren2004}.  Thus, odd-in-$B$ terms are a signature of magneto-chiral effects.

While non-linear transport is beyond the scope of the present work, we argue that the realistically-shaped 3DTI junction considered is an ideal platform to study magneto-chiral non-reciprocal transport.  Even if $z$-inversion symmetry is preserved as in Fig.~\ref{fig:constriction_schematic}, our system should act as a magneto-chiral non-reciprocal junction as soon as inversion symmetry in the $x$-$y$ plane orthogonal to transport is broken.  This is routinely done by using back- or top gates, whose symmetry breaking effect is particularly strong in 3DTIs.  In these materials the Dirac surface in proximity to the gate is strongly affected by it, and at the same time screens the one on the far side, which thus hardly feels the presence of the gate \cite{kozlov2016,ziegler2018,wang2020}.  As a consequence we expect a marked non-reciprocity especially in the low-energy regime of perfect chirality, when only chiral $m=0$ states are available for transport.

%

\section{Conclusion}
\label{sec_conclusions}

We studied the quantum magneto-transport properties of a realistically-shaped 3DTI nanowire.  The nanowire behaviour is mainly defined by the ratio between the length of the central narrow region $L_{center}$ and the magnetic length $l_B$, which actually exists only \textit{outside} such central region.  The shaped 3DTI junctions can thus be classified as \textit{long} -- $L_{center} \gg l_B$ -- or \textit{short} -- $L_{center} \lesssim l_B$.

Systems of realistic sizes were simulated via a 2D effective model based on a non-homogeneous grid.  Its validity was benchmarked and confirmed by comparison with full 3D simulations in scaled-down systems -- 3D systems being strongly limited in size due to numerical costs.

The essential features of the magneto-conductance traces are the result of: (i) Dirac particle in a box states, trapped in the central narrow region by side magnetic barriers; (ii) the appearance of quantum Hall states at the entrance/exit of the constriction.  The impact of the latter is particularly evident in short junctions and, irrespectively of the constriction length, at low energy.  Most importantly, in this regime the magnetic field forces a chiral behavior onto the junction, so that only states of a specific handedness are able to transport charge across it.  Such behavior can be understood via semiclassical arguments, validated by quantum mechanical numerical simulations.  We expect this tunable chirality to be exploitable for non-reciprocal transport in the non-linear regime.


\begin{acknowledgments}
This work was supported by Deutsche Forschungsgemeinschaft (DFG, German Research Foundation) within Project-ID 314695032-SFB 1277 (project A07).  CG thanks the University of Regensburg for its support and hospitality at various stages of this work. 
\end{acknowledgments}

\bibliographystyle{apsrev4-2}
\bibliography{biblio_Dirac_box}

\end{document}